\documentclass[aps,prl,twocolumn,showpacs,floatfix,superscriptaddress]{revtex4}

\usepackage{amsmath}
\usepackage{amssymb}
\usepackage{graphicx}

\begin{document}
\title{Unusual resonances in nanoplasmonic structures due to nonlocal response}
\date{\today}

\author{S\o ren Raza}
\affiliation{Department of Photonics Engineering, Technical University of Denmark, DK-2800 Kgs. Lyngby, Denmark}
\author{Giuseppe Toscano}
\affiliation{Department of Photonics Engineering, Technical University of Denmark, DK-2800 Kgs. Lyngby, Denmark}
\author{Antti-Pekka Jauho}
\affiliation{Department of Micro and Nanotechnology, Technical University of Denmark, DK-2800 Kgs. Lyngby, Denmark}
\author{Martijn Wubs}
\email{mwubs@fotonik.dtu.dk}
\affiliation{Department of Photonics Engineering, Technical University of Denmark, DK-2800 Kgs. Lyngby, Denmark}
\author{N. Asger Mortensen}
\email{asger@mailaps.org}
\affiliation{Department of Photonics Engineering, Technical University of Denmark, DK-2800 Kgs. Lyngby, Denmark}

\begin{abstract}
\noindent We study the nonlocal response of a confined electron gas
within the hydrodynamical Drude model. We address the question
whether plasmonic nanostructures exhibit nonlocal resonances that
have no counterpart in the local-response Drude model. Avoiding the
usual quasi-static approximation, we find that such resonances do
indeed occur, but only above the plasma frequency.
Thus the recently found nonlocal resonances at optical frequencies
for very small structures, obtained within quasi-static
approximation, are unphysical.
As a specific
example we consider nanosized metallic cylinders, for which extinction cross sections
and field distributions can be calculated analytically.

\end{abstract}


\maketitle

\noindent Nanoplasmonics~\cite{Gramotnev:2010a,Schuller:2010a} is
presently entering an exciting era where the metallic structures
offer nano-scale features that will eventually allow both photons
and electrons to exhibit their full wave nature. This regime challenges
the existing theoretical framework resting on a local-response
picture using bulk-material parameters. In tiny metallic
nanostructures, quantum
confinement~\cite{Lang:1970a,Zuloaga:2009a,PerezGonzalez:2010a,Ozturk:2011a,Wand:2011a}
and nonlocal
response~\cite{Bloch:1933a,Barton:1979a,Boardman:1982a,Pitarke:2007a,Abajo:2008a,VilloPerez:2009a,McMahon:2009a,McMahon:2010a,McMahon:2010b,McMahon:2010c,Toscano:2010a}
are believed to change the collective plasmonic behaviour with
resulting strong optical fingerprints and far-reaching consequences
for e.g. field-enhancement and extinction cross sections. Within
nonlocal response, Maxwell's constitutive relation between the
displacement and the electric fields reads
\begin{equation}
{\boldsymbol D}({\boldsymbol r},\omega) = \varepsilon_0 \int
d{\boldsymbol r}'\, {\boldsymbol\varepsilon}({\boldsymbol
r},{\boldsymbol r}',\omega) \cdot{\boldsymbol E} ({\boldsymbol
r}',\omega). \label{eq:realspace}
\end{equation}
The dielectric tensor $\boldsymbol \varepsilon({\boldsymbol
r},{\boldsymbol r}',\omega)$ reduces to $\varepsilon({\boldsymbol
r},\omega)\delta({\boldsymbol r}-{\boldsymbol r}')$ in the
local-response limit. Historically, there has been a strong emphasis
on nonlocal response in extended systems with translational
invariance (TI)~\cite{Boardman:1982a}, where a $k$-space representation is useful. However, for the
present problem of metallic nanostructures, TI is broken
and a real-space description is called for.

Recent theoretical studies of nanoscale plasmonic structures have
predicted considerable differences in the field distributions and
scattering cross sections between local and nonlocal response
theories, both in numerical implementations of a simplified
hydrodynamic Drude
model~\cite{McMahon:2009a,McMahon:2010a,McMahon:2010b,McMahon:2010c,Toscano:2010a},
and in corresponding  analytical calculations~\cite{McMahon:2010a}.
Importantly, new resonances of the free-electron plasma were found,
also at optical frequencies, which have no counterparts in
local-response theories.
Such novel resonances have already gained interest both from a fundamental~\cite{Wand:2011a} and an applied~\cite{Marty:2009a}  perspective.
At present, the status of these optical nonlocal resonances is unclear, since in Ref.~\onlinecite{VilloPerez:2009a}  the same nonlocal model was used as in Refs.~\onlinecite{McMahon:2009a,McMahon:2010a,McMahon:2010b,McMahon:2010c,Toscano:2010a},
and yet no corresponding new modes were found at visible frequencies.
%
Resolving this issue is important for the engineering of ultrasmall plasmonic structures with new functionalities~\cite{Marty:2009a,Xia:2009a,Peng:2010a}.

In this paper we report that unusual resonances due to nonlocal response do exist in nanoplasmonic structures,
but only above the plasma frequency, not in the visible.
We illustrate this property of arbitrary plasmonic structures by exact calculations for metallic cylinders.
We also clarify that different implementations of the common
quasi-static approximation~\cite{Barton:1979a,Pitarke:2007a}  are
the reason for  the conflicting results
in~\cite{McMahon:2009a,McMahon:2010a,McMahon:2010b,McMahon:2010c,Toscano:2010a} and~\cite{VilloPerez:2009a}.
Here we refrain from making this approximation altogether,
and by comparison analyze the validity and implementation of the quasi-static approximation in the hydrodynamic model.

\emph{The hydrodynamic Drude model.} We express the collective motion of
electrons in an inhomogeneous medium in terms of the electron
density $n({\boldsymbol r},t)$ and the hydrodynamical velocity
${\boldsymbol v}({\boldsymbol r},t)$ \cite{Bloch:1933a}. Under the
influence of macroscopic electromagnetic fields ${\boldsymbol
E}({\boldsymbol r},t)$ and ${\boldsymbol B}({\boldsymbol r},t)$, the
hydrodynamic model is defined via~\cite{Boardman:1982a}
\begin{equation}
\left[ \partial_t + {\boldsymbol v} \cdot {\boldsymbol \nabla} \right] {\boldsymbol v} = - \gamma {\boldsymbol v}
    -\frac{e}{m} \left[{\boldsymbol E} + {\boldsymbol v} \times {\boldsymbol B} \right] - \frac{ \beta^2}{n} {\boldsymbol \nabla} n, \label{eq:NLmotion}
\end{equation}
along with the continuity equation $\partial_t n = - {\boldsymbol \nabla} \cdot \left( n {\boldsymbol v} \right)$, expressing charge conservation.
In the right-hand side of Eq.~(\ref{eq:NLmotion}), the
$\gamma$-term represents damping, the second term is the Lorentz
force, while the third term is due to the internal kinetic energy of
the electron gas,  here described within the Thomas--Fermi model,
with $\beta$ proportional to the Fermi velocity $v_{\rm F}$. In
analogy with hydrodynamics, the third term represents a pressure
that gives rise to a nonlocal dielectric tensor, since energy may be
transported by other mechanisms than electromagnetic waves.

We follow the usual approach~\cite{Pitarke:2007a}
to solve Eq.~(\ref{eq:NLmotion}) and the continuity equation, by
expanding the physical fields in a zeroth-order static term, where
(e.g., $n_0$ is the homogeneous static electron density), and a
small (by assumption) first-order dynamic term, thereby linearizing
the equations. In the frequency domain, we obtain
\begin{subequations}
\label{eq:coupledequations}
\begin{equation}
    \beta^2 {\boldsymbol \nabla} \left[ {\boldsymbol \nabla} \cdot {\boldsymbol J} \right] + \omega\left(\omega+i\gamma\right) {\boldsymbol J} = i \omega \omega_{\rm p}^2 \varepsilon_0 {\boldsymbol E}, \label{eq:lmotion}
\end{equation}
for a homogeneous medium, where ${\boldsymbol J}({\boldsymbol r}) =
-en_0{\boldsymbol v}({\boldsymbol r})$ is the current density, and
$\omega_{\rm p}$ is the plasma frequency which also enters the Drude
local-response function $\varepsilon(\omega)=1-\omega_{\rm
p}^2/[\omega(\omega+i\gamma)]$. We focus on the plasma, leaving out
bulk interband effects present in real metals that could easily be
taken into account~\cite{Maier:2007a,McMahon:2009a}, as well as
band-bending effects at the metal surface.

\emph{The electromagnetic wave equation.} The retarded linearized hydrodynamic model is then fully described by Eq.~(\ref{eq:lmotion}) together with the Maxwell wave equation
\begin{equation}
{\boldsymbol\nabla}\times{\boldsymbol\nabla}\times{\boldsymbol E}=\frac{\omega^2}{c^2}{\boldsymbol E} + i\omega\mu_0 {\boldsymbol J}.\label{eq:Maxwell}
\end{equation}
\end{subequations}
In order to see that these coupled equations~(\ref{eq:coupledequations}) indeed describe nonlocal dielectric response,   one can in Eq.~(\ref{eq:Maxwell}) rewrite the current density ${\boldsymbol J}$ as an integral over the Green tensor of Eq.~(\ref{eq:lmotion}) and the electric field, whereby the nonlocal dielectric tensor of Eq.~(\ref{eq:realspace}) can be identified.

In a local-response description it is commonplace to introduce the
quasi-static or curl-free assumption that ${\boldsymbol \nabla}
\times {\boldsymbol E} = 0$~\cite{Jackson:1999a}. This
well-established approximation lies at the heart of most treatments
and interpretations of electromagnetic wave interactions with
sub-wavelength structures. Intuitively, one might expect that it can
be extended to the nonlocal case and indeed several nonlocal
treatments use this
assumption~\cite{Barton:1979a,Pitarke:2007a,VilloPerez:2009a,McMahon:2009a,McMahon:2010a,McMahon:2010b,McMahon:2010c}.
However, as we shall demonstrate, one should proceed with care.

\emph{Three models.} In this work we solve the
Eqs.~(\ref{eq:coupledequations}) directly, without further
assumptions or approximations. We also compare the {\em nonlocal
model} with two other models obtained by further assumptions. The
{\em curl-free nonlocal model} enforces the condition ${\boldsymbol
\nabla} \times {\boldsymbol E}=0$, which with Eq.~(\ref{eq:lmotion})
implies that also ${\boldsymbol\nabla} \times {\boldsymbol J} =0$ in
the medium.
For the differential-operator term in Eq.~(\ref{eq:lmotion}), from
now on denoted $\hat{L}_J$, this has the consequence that
${\boldsymbol \nabla} [{\boldsymbol \nabla} \cdot ]$ simplifies to
the Laplace operator ${\boldsymbol \nabla}^2$, which gives the model
used by Ruppin in the context of exciton physics
in~\cite{Ruppin:1989a}, and recently in plasmonics by
McMahon~\emph{et
al.}~\cite{McMahon:2009a,McMahon:2010a,McMahon:2010b,McMahon:2010c}
and also by ourselves~\cite{Toscano:2010a}. Finally, by assuming
$\hat{L}_J= 0$ in the hydrodynamic treatment~(\ref{eq:lmotion}), the
familiar {\em local model} is obtained, with $\boldsymbol J$ and
$\boldsymbol E$ related by Ohm's law.

We assume that the static density of electrons $n_0$ vanishes
outside the metal of volume $V$, while it is constant and equal to
the bulk value inside $V$, thus neglecting tunneling effects and
inhomogeneous electron distributions associated with quantum
confinement~\cite{Lang:1970a,Ozturk:2011a}. As a consequence,
${\boldsymbol J}= 0$ outside $V$
for all three models.

\begin{table}[b]
\begin{tabular}{cc|c|c|c|c|c|}
\cline{3-7}
 &&\multicolumn{2}{|c|}{${\boldsymbol r}\in V$} & \multicolumn{2}{|c|}{${\boldsymbol r}\in \partial V$}& ${\boldsymbol r}\in\hspace{-2.6mm}/\hspace{1.8mm} V$\\\cline{3-7}
&& ${\boldsymbol \nabla}\times {\boldsymbol J}$ &$\hat{L}_J$ &$\hat{\boldsymbol n}\cdot {\boldsymbol J}$ & $\hat{\boldsymbol n}\times {\boldsymbol J}$&${\boldsymbol J}$\\\hline
\multicolumn{2}{|c|}{local}& $\neq 0$ &$ 0$ & $0$ & $\neq 0$& $0$\\
\multicolumn{2}{|c|}{nonlocal} &  $\neq 0$ & $\beta^{2}{\boldsymbol \nabla}[{\boldsymbol \nabla\cdot}]$& $0$ & $\neq 0$ & $0$\\
\multicolumn{2}{|c|}{nonlocal (curl-free)
}
&$ 0$& $\beta^{2} {\boldsymbol \nabla}^2$ & $0$ & $0$& $0$\\
\hline
\end{tabular}
\caption{Summary of the three different response models.
$V$ is the volume of the nanostructure, and $\partial V$ its boundary.}
\label{Table1}
\end{table}

\emph{Boundary conditions.} In the local model the current ${\boldsymbol J}$
has the same the spatial dependence as the ${\boldsymbol E}$--field.
Thus, in this case there
are no additional boundary conditions (ABCs) to those already used in
Maxwell's equations. For the nonlocal-response models on the other
hand, ABCs are in general
needed~\cite{Sauter:1967a,Melnyk:1970a,Jewsbury:1981a,Boardman:1982a,McMahon:2010b}.
From discussions in the literature it might appear that the number
of necessary ABCs is a subtle issue, but we emphasize that there
should be no ambiguity. The crucial point is that the required
number of ABCs depends on the assumed static electron density
profile at the boundaries~\cite{Jewsbury:1981a}.
For the present problem with the electron density vanishing identically outside the metal, only one ABC is needed in the nonlocal model to obtain unique solutions~\cite{Jewsbury:1981a}, and it is readily found from the continuity equation and Gauss' theorem: $\hat{\boldsymbol n}\cdot{\boldsymbol J}=0$ on the boundary, where $\hat{\boldsymbol n}$ is a normal vector to the surface, i.e. the normal-component of the current vanishes
~\cite{Sauter:1967a,Jewsbury:1981a,Boardman:1982a}, for all three models.
On the other hand, in general the tangential current $\hat{\boldsymbol n}\times{\boldsymbol J}$ is
non-zero. This `slip' of the current is not surprising, since the hydrodynamic equation~(\ref{eq:NLmotion}) describes the plasma as a non-viscous fluid.

Likewise, in several implementations of the quasi-static approximation, no further ABCs are needed to uniquely determine the electric field and current density~\cite{Pitarke:2007a,VilloPerez:2009a}.
In contrast, in the curl-free nonlocal model of Refs.~\cite{Ruppin:1989a,McMahon:2009a,McMahon:2010a,McMahon:2010b,McMahon:2010c,Toscano:2010a}
one more ABC {\em is} needed. Assumed is that the tangential components of ${\boldsymbol J}$ vanish at the boundary ($\hat{\boldsymbol n}\times{\boldsymbol J}=0$), so that both normal and tangential components of the current field
vanish on the boundary. In the different context of exciton physics~\cite{Ruppin:1989a} these are often referred to as Pekar's
additional boundary conditions. There, the vanishing of the tangential boundary currents
is motivated by the physical assumption that exciton wave functions vanish on the boundary~\cite{Pekar:1958a,Ruppin:1989a}.
Instead, in the hydrodynamical theory of metals, the ABC $\hat{\boldsymbol n}\times{\boldsymbol J}=0$ seems more {\em ad hoc}: not a direct consequence of the quasi-static approximation, and not correct if that  approximation is not made.
 The different boundary conditions are summarized in Table~\ref{Table1}.

\emph{Extinction cross section of metallic nanowires.} To illustrate
the surprisingly different physical consequences of the three
models, we consider light scattering by a nanowire. Rather than
solving Eqs.~(\ref{eq:coupledequations}) numerically for a general
cross-sectional geometry, we here limit our analysis to cylindrical
wires whereby significant analytical progress is possible. We use an
extended Mie theory, developed by
Ruppin~\cite{Ruppin:1989a,Ruppin:2001a}, to calculate the extinction
cross section $\sigma_\text{ext}$ of an infinitely long spatially
dispersive cylindrical metal nanowire in vacuum.  Outside the wire
there are incoming and scattered fields (both divergence-free),
whereas inside the wire both divergence-free and curl-free modes can
be excited, the latter type only in case of nonlocal response. The
cross section is~\cite{Hulst:1957a}
\begin{equation}\label{sigma_ext}
    \sigma_\text{ext} = -\frac{2}{k_0 a} \sum_{n=-\infty}^{\infty} {\rm Re}\{a_n\},
\end{equation}
where $a$ is the radius, $k_0=\omega/c$ is the vacuum wave
vector, and $a_n$ is a cylindrical Bessel-function expansion
coefficient for the scattered fields. We consider a normally
incident plane wave with the electric-field polarization
perpendicular to the cylinder axis (TM).
The expression for the coefficients $a_n$ depends on the particular response model and the associated ABCs.
For the curl-free nonlocal model, the $a_n$ are
known~\cite{Ruppin:1989a}. For the full hydrodynamic model we follow
the approach of Ref.~\onlinecite{Ruppin:2001a}, where the ABC of
Ref.~\onlinecite{Melnyk:1970a} is employed. This ABC is for metals in free
space equivalent to $\hat{\boldsymbol n}\cdot{\boldsymbol J}=0$.
We obtain
\begin{equation}
    a_n =  - \frac{\left[ d_n + J_n'(\kappa_t a) \right]J_n(k_0a)  -  \sqrt{\varepsilon} J_n(\kappa_t a) J_n'(k_0a)} {\left[d_n +  J_n'(\kappa_t a)\right]H_n(k_0a) - \sqrt{\varepsilon} J_n(\kappa_t a) H_n'(k_0a)}, \label{eq:an_hyd}
\end{equation}
where $J_n$ and $H_n$ are Bessel and Hankel functions of the first
kind and $\kappa_t^2=\varepsilon(\omega) k_0^2$. The $d_n$ coefficients are
\begin{equation}
    d_n = \frac{n^2}{\kappa_l a} \frac{J_n(\kappa_l a)}{J_n'(\kappa_la)} \frac{J_n(\kappa_ta)}{\kappa_ta} \left[\varepsilon(\omega) - 1\right],
\end{equation}
where $\kappa_l^2=(\omega^2+i\omega\gamma-\omega_{\rm p}^2)/\beta^2$.
In the limit $\beta \rightarrow 0$, the
$d_n$ vanish and the $a_n$ of Eq.~(\ref{eq:an_hyd}) reduce to the local Drude scattering coefficients~\cite{Hulst:1957a}, which confirms that nonlocal response in our model requires moving charges.

\begin{figure}[t]
\includegraphics[width=0.5\textwidth]{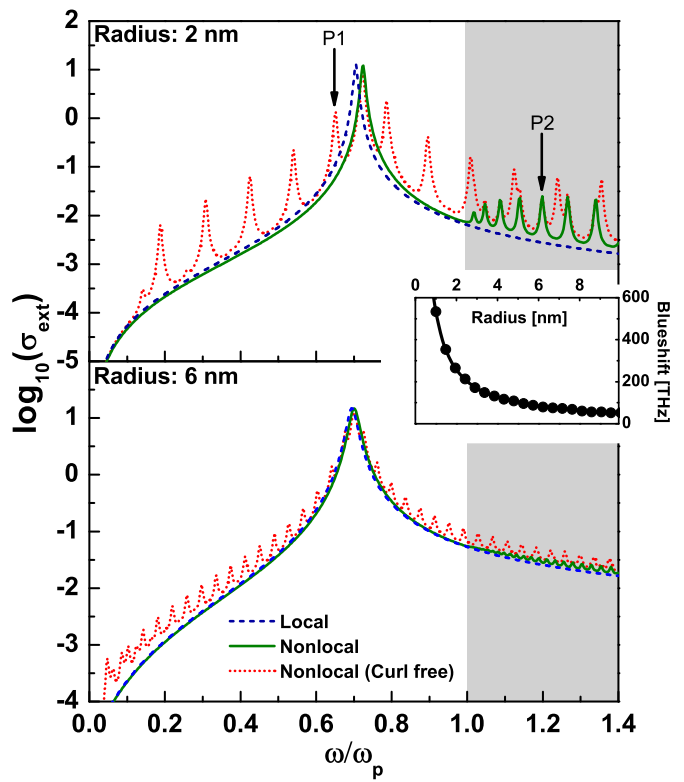}
\caption{(Color online) Extinction cross sections $\sigma_{\rm ext}$ as a function of frequency for  TM-polarized light  normally incident on a metallic cylinder in vacuum.
Parameters for Au as in Ref.~\onlinecite{McMahon:2009a}: $\hbar \omega_{\rm p} = 8.812\,{\rm eV}$, $\hbar\gamma = 0.0752\,{\rm eV}$, and $v_{\rm F} = 1.39 * 10^{6}\,{\rm m/s}$.
 Inset:
 frequency shift of the maximum $\sigma_{\rm ext}(\omega)$ for nonlocal against local response,  as a function of
 radius.
 }
\label{fig:crosssection}
\end{figure}

\emph{Are there nonlocal resonances?} Figure~\ref{fig:crosssection}
depicts the extinction cross section of Eq.~(\ref{sigma_ext}) for two cylinder radii,
comparing the nonlocal models with the local Drude model.
The main surface-plasmon resonance peak at $\omega_{\rm p}/\sqrt{2}$ is blueshifted
as compared to the local model, and more so for smaller radii.
Similar blueshifts have been reported for other
geometries~\cite{Abajo:2008a} and in the curl-free nonlocal
model~\cite{Ruppin:1989a,McMahon:2009a}.

Figure 1 shows the unusual resonances mentioned in the title of this paper:
additional peaks {\it do} appear in the nonlocal theory
but only for frequencies {\em above} the plasma frequency
$\omega_{\rm p}$ ($\hbar \omega_{\rm p}=8.9\,{\rm eV}$ for Ag and
Au; $1.5$ to $3\,{\rm eV}$ is visible). These peaks (such as P2 in
Fig.~\ref{fig:crosssection}) are due to the excitation of confined
longitudinal modes, which are bulk-plasmon states with discrete
energies above $\hbar \omega_{\rm p}$ due to confinement in the
cylinder~\cite{VilloPerez:2009a}. These peaks are analogous to
discrete absorption lines above the band gap in quantum-confined
semiconductor structures. Interestingly, contrary to the common
belief that
light does not scatter off bulk plasmons,
which is correct in the local theory ({\em i.e.} no peak around
$\omega_{\rm p}$ in Fig.~\ref{fig:crosssection}), here in the
nonlocal model we do find such a coupling to longitudinal modes. The
new resonances could therefore be observed with electron loss
spectroscopy but also with extreme UV light. The curl-free model
also exhibits these resonances.

The striking difference between the two nonlocal-response models is
that the curl-free nonlocal model shows additional stronger
resonances, both above and below the plasma frequency, such as P1 in
Fig.~\ref{fig:crosssection}, in particular also at optical
frequencies. These peaks do not show up in the full hydrodynamical
model,
and thus originate from a mathematical approximation rather than a physical
mechanism.
It would however be premature to conclude that the quasi-static
approximation
breaks down, because in
Ref.~\onlinecite{VilloPerez:2009a} the modes of cylinders in the
hydrodynamical Drude model were found after making the quasi-static
approximation, and the only novel modes found were the confined bulk
plasmon modes above $\omega_{\rm p}$.
Fig.~\ref{fig:crosssection} also illustrates that for increasing
radii, $\sigma_{\rm ext}$ in  the two nonlocal models converges
towards the local-response value.
This convergence is slower for the curl-free model.

\begin{figure}[t]
\includegraphics[width=0.5\textwidth]{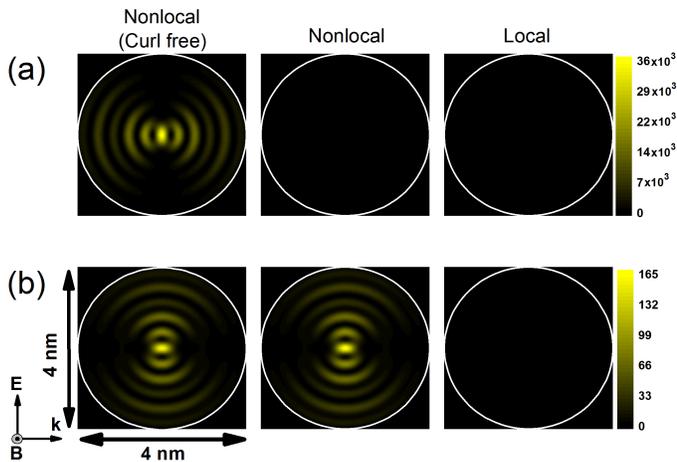}
\caption{(Color online) Field distributions in the three different models, for  TM-polarized light normally incident on a cylinder of radius $a=2$\,nm. (a) Normalized displacement field $|{\boldsymbol D}|^2/|{\boldsymbol D}_\text{in}|^2$ at the frequency $\omega=0.6503\omega_{\rm p}$ (P1 in Fig.~\ref{fig:crosssection}). ${\boldsymbol D}_\text{in}=\varepsilon_0 {\boldsymbol E}_\text{in}$ and ${\boldsymbol E}_\text{in}$ is the incident electric field. (b) Analogous plots of $|{\boldsymbol E}|^2/|{\boldsymbol E}_\text{in}|^2$ for $\omega=1.1963\omega_{\rm p}$ (P2 in Fig.~\ref{fig:crosssection}).}
\label{fig:Field_plots}
\end{figure}

In Figure~\ref{fig:Field_plots}(a)
we depict the scaled {\em displacement}-field distributions for the
three models at the frequency marked P1 in Fig. 1, where only the
curl-free nonlocal model has a (spurious) resonance.
Correspondingly, in Fig.~\ref{fig:Field_plots}(a) we find a
standing-wave pattern only in that model. Its appearance in the
displacement field illustrates that the spurious resonance is a
transverse resonance, {\em i.e.} occurring in the divergence-free
components of ${\boldsymbol E}$ and ${\boldsymbol J}$.
Fig.~\ref{fig:Field_plots}(b) on the other hand shows the normalized
{\em electric}-field intensity for a true resonant mode at the
frequency P2 of Fig.~\ref{fig:crosssection}. Only the two nonlocal
models give rise to  resonant electric-field patterns. These
confined bulk plasmon modes are longitudinal and would not produce
standing waves in the displacement field.

\emph{Origin of spurious resonances.}  By eliminating the electric
field from  Eqs.~(\ref{eq:coupledequations}), it follows that the
exact hydrodynamic current satisfies the pair of third-order
equations
\begin{subequations}
\label{eq:generalizedJ}
\begin{eqnarray}
\left(\beta^2\nabla^2+\omega^2+i\omega\gamma-\omega_{\rm p}^2\right){\boldsymbol \nabla}\cdot{\boldsymbol J}&=&0 \label{BoardmandivJ}\\
\left(c^2\nabla^2+\omega^2\varepsilon(\omega)\right){\boldsymbol \nabla}\times{\boldsymbol J}&=&0, \label{BoardmancurlJ}
\end{eqnarray}
\end{subequations}
which reduce to the more symmetric Boardman
equations~\cite{Boardman:1977a} in the absence of damping.
For arbitrary geometry,
Eq.~(\ref{BoardmandivJ}) has damped solutions of  ${\boldsymbol
\nabla}\cdot{\boldsymbol J}$ for $\omega <\omega_{\rm p}$ and
finite-width resonances for $\omega > \omega_{\rm p}$, as seen in  Fig.~\ref{fig:crosssection}.
Both solutions can be consistent
with the quasi-static approximation ${\boldsymbol
\nabla}\times{\boldsymbol J}=0$ that trivially solves
Eq.~(\ref{BoardmancurlJ}).
On the other hand, we find that the spurious resonances have
resonant
divergence-free components of ${\bf E}$ and $\bf J$. However, these
cannot at the same time be curl-free. Thus the curl-free nonlocal model has resonant solutions
with nonvanishing curl, which is logically inconsistent. But how
could this arise? Once the ${\boldsymbol \nabla}\times{\boldsymbol
J} = 0$ assumption has been invoked to simplify the differential
operator into $\hat{L}_J=\beta^{2} \nabla^2$, the resulting
Laplacian equation analogous to (\ref{eq:lmotion}) carries
no information that the resulting solution should also
be curl-free. Thus, the solutions found for this equation are not
necessarily self-consistent.

\emph{Conclusions.} We have shown that plasmonic nanostructures exhibit novel resonances due to
nonlocal response in the hydrodynamic Drude
model, but only above the plasma frequency. The recently
reported nonlocal resonances in the
visible~\cite{McMahon:2009a,McMahon:2010a,McMahon:2010b,McMahon:2010c,Toscano:2010a} agree with older work~\cite{Ruppin:1989a}, but
are a surprisingly pronounced consequence of an implementation of
the quasi-static approximation that is not self-consistent. For
nanowires, we find extinction resonances without making the
quasi-static approximation that agree with the quasi-static modes of
Ref.~\onlinecite{VilloPerez:2009a}, so we do not claim a general breakdown
of the
approximation itself. Even though there are no nonlocal resonances
in the visible, plasmonic field enhancements are affected by
nonlocal response. For arbitrary geometries, numerical methods must
be used to quantitatively assess their importance. Self-consistent
versions of the versatile
time-domain~\cite{McMahon:2009a,McMahon:2010a,McMahon:2010b,McMahon:2010c}
and frequency-domain~\cite{Toscano:2010a}
implementations of the hydrodynamical model  can
do just that.

\begin{acknowledgments}
This work was financially supported by Danish Research
Council for Technology and Production Sciences (Grant No. 274-07-0080),
and by the FiDiPro program of the Finnish Academy.
\end{acknowledgments}



\begin{thebibliography}{31}
\expandafter\ifx\csname natexlab\endcsname\relax\def\natexlab#1{#1}\fi
\expandafter\ifx\csname bibnamefont\endcsname\relax
  \def\bibnamefont#1{#1}\fi
\expandafter\ifx\csname bibfnamefont\endcsname\relax
  \def\bibfnamefont#1{#1}\fi
\expandafter\ifx\csname citenamefont\endcsname\relax
  \def\citenamefont#1{#1}\fi
\expandafter\ifx\csname url\endcsname\relax
  \def\url#1{\texttt{#1}}\fi
\expandafter\ifx\csname urlprefix\endcsname\relax\def\urlprefix{URL }\fi
\providecommand{\bibinfo}[2]{#2}
\providecommand{\eprint}[2][]{\url{#2}}

\bibitem[{\citenamefont{Gramotnev and Bozhevolnyi}(2010)}]{Gramotnev:2010a}
\bibinfo{author}{\bibfnamefont{D.~K.} \bibnamefont{Gramotnev}}
  \bibnamefont{and} \bibinfo{author}{\bibfnamefont{S.~I.}
  \bibnamefont{Bozhevolnyi}}, \bibinfo{journal}{Nat. Photon.}
  \textbf{\bibinfo{volume}{4}}, \bibinfo{pages}{83} (\bibinfo{year}{2010}).

\bibitem[{
\citenamefont{Schuller, Barnard, Cai, Jun, White, and Brongersma}
}]{Schuller:2010a}
\bibinfo{author}{\bibfnamefont{J.~A.} \bibnamefont{Schuller}},
  \bibinfo{author}{\bibfnamefont{E.~S.} \bibnamefont{Barnard}},
  \bibinfo{author}{\bibfnamefont{W.}~\bibnamefont{Cai}},
  \bibinfo{author}{\bibfnamefont{Y.~C.} \bibnamefont{Jun}},
  \bibinfo{author}{\bibfnamefont{J.~S.} \bibnamefont{White}}, \bibnamefont{and}
  \bibinfo{author}{\bibfnamefont{M.~L.} \bibnamefont{Brongersma}},
  \bibinfo{journal}{Nat. Mater.} \textbf{\bibinfo{volume}{9}},
  \bibinfo{pages}{193} (\bibinfo{year}{2010}).

\bibitem[{\citenamefont{Lang and Kohn}(1970)}]{Lang:1970a}
\bibinfo{author}{\bibfnamefont{N.~D.} \bibnamefont{Lang}} \bibnamefont{and}
  \bibinfo{author}{\bibfnamefont{W.}~\bibnamefont{Kohn}},
  \bibinfo{journal}{Phys. Rev. B} \textbf{\bibinfo{volume}{1}},
  \bibinfo{pages}{4555} (\bibinfo{year}{1970}).

\bibitem[{
\citenamefont{Zuloaga, Prodan, and
  Nordlander}}]{Zuloaga:2009a}
\bibinfo{author}{\bibfnamefont{J.}~\bibnamefont{Zuloaga}},
  \bibinfo{author}{\bibfnamefont{E.}~\bibnamefont{Prodan}}, \bibnamefont{and}
  \bibinfo{author}{\bibfnamefont{P.}~\bibnamefont{Nordlander}},
  \bibinfo{journal}{Nano Lett.} \textbf{\bibinfo{volume}{9}},
  \bibinfo{pages}{887 } (\bibinfo{year}{2009}).

\bibitem[{
  \citenamefont{P{\'e}rez-Gonz{\'a}lez, Zabala, Borisov, Halas,
  Nordlander, and Aizpurua}}]{PerezGonzalez:2010a}
\bibinfo{author}{\bibfnamefont{O.}~\bibnamefont{P{\'e}rez-Gonz{\'a}lez}},
  \bibinfo{author}{\bibfnamefont{N.}~\bibnamefont{Zabala}},
  \bibinfo{author}{\bibfnamefont{A.~G.} \bibnamefont{Borisov}},
  \bibinfo{author}{\bibfnamefont{N.~J.} \bibnamefont{Halas}},
  \bibinfo{author}{\bibfnamefont{P.}~\bibnamefont{Nordlander}},
  \bibnamefont{and} \bibinfo{author}{\bibfnamefont{J.}~\bibnamefont{Aizpurua}},
  \bibinfo{journal}{Nano Lett.} \textbf{\bibinfo{volume}{10}},
  \bibinfo{pages}{3090 } (\bibinfo{year}{2010}).

\bibitem[{
\citenamefont{{\"Ozt\"urk},
  Xiao, Yan, Wubs, Jauho, and Mortensen}}]{Ozturk:2011a}
\bibinfo{author}{\bibfnamefont{Z.~F.} \bibnamefont{{\"Ozt\"urk}}},
  \bibinfo{author}{\bibfnamefont{S.}~\bibnamefont{Xiao}},
  \bibinfo{author}{\bibfnamefont{M.}~\bibnamefont{Yan}},
  \bibinfo{author}{\bibfnamefont{M.}~\bibnamefont{Wubs}},
  \bibinfo{author}{\bibfnamefont{A.-P.} \bibnamefont{Jauho}}, \bibnamefont{and}
  \bibinfo{author}{\bibfnamefont{N.~A.} \bibnamefont{Mortensen}},
  \bibinfo{journal}{J. Nanophoton.} \textbf{\bibinfo{volume}{5}},
  \bibinfo{pages}{051602} (\bibinfo{year}{2011}).

\bibitem[{
\citenamefont{Wand, Schindlmayer,
  Meier, and F{\" o}rstner}}]{Wand:2011a}
\bibinfo{author}{\bibfnamefont{M.}~\bibnamefont{Wand}},
  \bibinfo{author}{\bibfnamefont{A.}~\bibnamefont{Schindlmayer}},
  \bibinfo{author}{\bibfnamefont{T.}~\bibnamefont{Meier}}, \bibnamefont{and}
  \bibinfo{author}{\bibfnamefont{J.}~\bibnamefont{F{\" o}rstner}},
  \bibinfo{journal}{Phys. Status Solidi B} \textbf{\bibinfo{volume}{248}},
  \bibinfo{pages}{887} (\bibinfo{year}{2011}).

\bibitem[{\citenamefont{Bloch}(1933)}]{Bloch:1933a}
\bibinfo{author}{\bibfnamefont{F.}~\bibnamefont{Bloch}},
  \bibinfo{journal}{Zeitschrift f\text{\"{u}}r Physik A 
  }
  \textbf{\bibinfo{volume}{81}}, \bibinfo{pages}{363} (\bibinfo{year}{1933}).

\bibitem[{\citenamefont{Barton}(1979)}]{Barton:1979a}
\bibinfo{author}{\bibfnamefont{G.}~\bibnamefont{Barton}},
  \bibinfo{journal}{Rep. Prog. Phys.} \textbf{\bibinfo{volume}{42}},
  \bibinfo{pages}{963} (\bibinfo{year}{1979}).

\bibitem[{\citenamefont{Boardman}(1982)}]{Boardman:1982a}
\bibinfo{author}{\bibfnamefont{A.~D.} \bibnamefont{Boardman}},
  \emph{\bibinfo{title}{Electromagnetic Surface Modes.
  }} (\bibinfo{publisher}{Wiley}, \bibinfo{address}{New York}, \bibinfo{year}{1982}).

\bibitem[{
\citenamefont{Pitarke, Silkin,
  Chulkov, and Echenique}}]{Pitarke:2007a}
\bibinfo{author}{\bibfnamefont{J.}~\bibnamefont{Pitarke}},
  \bibinfo{author}{\bibfnamefont{V.}~\bibnamefont{Silkin}},
  \bibinfo{author}{\bibfnamefont{E.}~\bibnamefont{Chulkov}}, \bibnamefont{and}
  \bibinfo{author}{\bibfnamefont{P.}~\bibnamefont{Echenique}},
  \bibinfo{journal}{Rep. Prog. Phys.} \textbf{\bibinfo{volume}{70}},
  \bibinfo{pages}{1} (\bibinfo{year}{2007}).

\bibitem[{\citenamefont{Garc{\'i}a~de Abajo}(2008)}]{Abajo:2008a}
\bibinfo{author}{\bibfnamefont{F.~J.} \bibnamefont{Garc{\'i}a~de Abajo}},
  \bibinfo{journal}{J. Phys. Chem. C} \textbf{\bibinfo{volume}{112}},
  \bibinfo{pages}{17983} (\bibinfo{year}{2008}).

\bibitem[{\citenamefont{Vill{\' o}-P{\'e}rez and
  Arista}(2009)}]{VilloPerez:2009a}
\bibinfo{author}{\bibfnamefont{I.}~\bibnamefont{Vill{\' o}-P{\'e}rez}}
  \bibnamefont{and} \bibinfo{author}{\bibfnamefont{N.~R.}
  \bibnamefont{Arista}}, \bibinfo{journal}{Surf. Sci.}
  \textbf{\bibinfo{volume}{603}}, \bibinfo{pages}{1} (\bibinfo{year}{2009}).

\bibitem[{
\citenamefont{McMahon, Gray, and
  Schatz}}]{McMahon:2009a}
\bibinfo{author}{\bibfnamefont{J.~M.} \bibnamefont{McMahon}},
  \bibinfo{author}{\bibfnamefont{S.~K.} \bibnamefont{Gray}}, \bibnamefont{and}
  \bibinfo{author}{\bibfnamefont{G.~C.} \bibnamefont{Schatz}},
  \bibinfo{journal}{Phys. Rev. Lett.} \textbf{\bibinfo{volume}{103}},
  \bibinfo{pages}{097403} (\bibinfo{year}{2009}).

\bibitem[{
\citenamefont{McMahon, Gray, and
  Schatz}}]{McMahon:2010a}
\bibinfo{author}{\bibfnamefont{J.~M.} \bibnamefont{McMahon}},
  \bibinfo{author}{\bibfnamefont{S.~K.} \bibnamefont{Gray}}, \bibnamefont{and}
  \bibinfo{author}{\bibfnamefont{G.~C.} \bibnamefont{Schatz}},
  \bibinfo{journal}{Phys. Rev. B} \textbf{\bibinfo{volume}{82}},
  \bibinfo{pages}{035423} (\bibinfo{year}{2010}{\natexlab{a}}).

\bibitem[{
\citenamefont{McMahon, Gray, and
  Schatz}}]{McMahon:2010b}
\bibinfo{author}{\bibfnamefont{J.~M.} \bibnamefont{McMahon}},
  \bibinfo{author}{\bibfnamefont{S.~K.} \bibnamefont{Gray}}, \bibnamefont{and}
  \bibinfo{author}{\bibfnamefont{G.~C.} \bibnamefont{Schatz}},
  \bibinfo{journal}{Nano Lett.} \textbf{\bibinfo{volume}{10}},
  \bibinfo{pages}{3473} (\bibinfo{year}{2010}{\natexlab{b}}).

\bibitem[{\citenamefont{McMahon et~al.}(2010{\natexlab{c}})
\citenamefont{McMahon, Gray, and Schatz}}]{McMahon:2010c}
\bibinfo{author}{\bibfnamefont{J.~M.} \bibnamefont{McMahon}},
  \bibinfo{author}{\bibfnamefont{S.~K.} \bibnamefont{Gray}}, \bibnamefont{and}
  \bibinfo{author}{\bibfnamefont{G.~C.} \bibnamefont{Schatz}},
  \bibinfo{journal}{J. Phys. Chem. C} \textbf{\bibinfo{volume}{114}},
  \bibinfo{pages}{15903} (\bibinfo{year}{2010}{\natexlab{c}}).

\bibitem[{
\citenamefont{Toscano, Wubs, Xiao,
  Yan, {\"Ozt\"urk}, Jauho, and Mortensen}}]{Toscano:2010a}
\bibinfo{author}{\bibfnamefont{G.}~\bibnamefont{Toscano}},
  \bibinfo{author}{\bibfnamefont{M.}~\bibnamefont{Wubs}},
  \bibinfo{author}{\bibfnamefont{S.}~\bibnamefont{Xiao}},
  \bibinfo{author}{\bibfnamefont{M.}~\bibnamefont{Yan}},
  \bibinfo{author}{\bibfnamefont{Z.~F.} \bibnamefont{{\"Ozt\"urk}}},
  \bibinfo{author}{\bibfnamefont{A.-P.} \bibnamefont{Jauho}}, \bibnamefont{and}
  \bibinfo{author}{\bibfnamefont{N.~A.} \bibnamefont{Mortensen}},
  \bibinfo{journal}{Proc. SPIE} \textbf{\bibinfo{volume}{7757}},
  \bibinfo{pages}{77571T} (\bibinfo{year}{2010}).

\bibitem[{
\citenamefont{Marty, Arbouet,
  Girard, Margueritat, Gonzalo, and Afonso}}]{Marty:2009a}
\bibinfo{author}{\bibfnamefont{R.}~\bibnamefont{Marty}},
  \bibinfo{author}{\bibfnamefont{A.}~\bibnamefont{Arbouet}},
  \bibinfo{author}{\bibfnamefont{C.}~\bibnamefont{Girard}},
  \bibinfo{author}{\bibfnamefont{J.}~\bibnamefont{Margueritat}},
  \bibinfo{author}{\bibfnamefont{J.}~\bibnamefont{Gonzalo}}, \bibnamefont{and}
  \bibinfo{author}{\bibfnamefont{C.~N.} \bibnamefont{Afonso}},
  \bibinfo{journal}{J. Chem. Phys.} \textbf{\bibinfo{volume}{131}},
  \bibinfo{pages}{224707} (\bibinfo{year}{2009}).

\bibitem[{
\citenamefont{Xia, Xiong, Lim, and
  Skrabalak}}]{Xia:2009a}
\bibinfo{author}{\bibfnamefont{Y.}~\bibnamefont{Xia}},
  \bibinfo{author}{\bibfnamefont{Y.}~\bibnamefont{Xiong}},
  \bibinfo{author}{\bibfnamefont{B.}~\bibnamefont{Lim}}, \bibnamefont{and}
  \bibinfo{author}{\bibfnamefont{S.~E.} \bibnamefont{Skrabalak}},
  \bibinfo{journal}{Angew. Chem. Int. Ed.} \textbf{\bibinfo{volume}{48}},
  \bibinfo{pages}{60} (\bibinfo{year}{2009}).

\bibitem[{
\citenamefont{Peng, McMahon, Schatz,
  Gray, and Sun}}]{Peng:2010a}
\bibinfo{author}{\bibfnamefont{S.}~\bibnamefont{Peng}},
  \bibinfo{author}{\bibfnamefont{J.~M.} \bibnamefont{McMahon}},
  \bibinfo{author}{\bibfnamefont{G.~C.} \bibnamefont{Schatz}},
  \bibinfo{author}{\bibfnamefont{S.~K.} \bibnamefont{Gray}}, \bibnamefont{and}
  \bibinfo{author}{\bibfnamefont{Y.}~\bibnamefont{Sun}},
  \bibinfo{journal}{Proc. Natl. Acad. Sci. U.S.A.} \textbf{\bibinfo{volume}{107}},
  \bibinfo{pages}{14530} (\bibinfo{year}{2010}).

\bibitem[{\citenamefont{Maier}(2007)}]{Maier:2007a}
\bibinfo{author}{\bibfnamefont{S.~A.} \bibnamefont{Maier}},
  \emph{\bibinfo{title}{Plasmonics: Fundamentals and Applications}}
  (\bibinfo{publisher}{Springer}, \bibinfo{address}{New York},
  \bibinfo{year}{2007}).

\bibitem[{\citenamefont{Jackson}(1999)}]{Jackson:1999a}
\bibinfo{author}{\bibfnamefont{J.~D.} \bibnamefont{Jackson}},
  \emph{\bibinfo{title}{Classical Electrodynamics}}, \bibinfo{edition}{{3rd} ed.} (\bibinfo{publisher}{Wiley}, \bibinfo{address}{Hoboken, NJ}, \bibinfo{year}{1999})


\bibitem[{\citenamefont{Ruppin}(1989)}]{Ruppin:1989a}
\bibinfo{author}{\bibfnamefont{R.}~\bibnamefont{Ruppin}}, \bibinfo{journal}{J.
  Opt. Soc. Am. B} \textbf{\bibinfo{volume}{6}}, \bibinfo{pages}{1559}
  (\bibinfo{year}{1989}).

\bibitem[{\citenamefont{Sauter}(1967)}]{Sauter:1967a}
\bibinfo{author}{\bibfnamefont{F.}~\bibnamefont{Sauter}}, \bibinfo{journal}{Z.
  Phys.} \textbf{\bibinfo{volume}{203}}, \bibinfo{pages}{488}
  (\bibinfo{year}{1967}).

\bibitem[{\citenamefont{Melnyk and Harrison}(1970)}]{Melnyk:1970a}
\bibinfo{author}{\bibfnamefont{A.~R.} \bibnamefont{Melnyk}} \bibnamefont{and}
  \bibinfo{author}{\bibfnamefont{M.~J.} \bibnamefont{Harrison}},
  \bibinfo{journal}{Phys. Rev. B} \textbf{\bibinfo{volume}{2}},
  \bibinfo{pages}{835} (\bibinfo{year}{1970}).

\bibitem[{\citenamefont{Jewsbury}(1981)}]{Jewsbury:1981a}
\bibinfo{author}{\bibfnamefont{P.}~\bibnamefont{Jewsbury}},
  \bibinfo{journal}{J. Phys. F: Metal Phys.} \textbf{\bibinfo{volume}{11}},
  \bibinfo{pages}{195} (\bibinfo{year}{1981}).

\bibitem[{\citenamefont{Pekar}(1958)}]{Pekar:1958a}
\bibinfo{author}{\bibfnamefont{S.}~\bibnamefont{Pekar}}, \bibinfo{journal}{J.
  Phys. Chem. Solids} \textbf{\bibinfo{volume}{5}}, \bibinfo{pages}{11}
  (\bibinfo{year}{1958}).

\bibitem[{\citenamefont{Ruppin}(2001)}]{Ruppin:2001a}
\bibinfo{author}{\bibfnamefont{R.}~\bibnamefont{Ruppin}},
  \bibinfo{journal}{Opt. Commun.} \textbf{\bibinfo{volume}{190}},
  \bibinfo{pages}{205} (\bibinfo{year}{2001}).

\bibitem[{\citenamefont{van~de Hulst}(1957)}]{Hulst:1957a}
\bibinfo{author}{\bibfnamefont{H.}~\bibnamefont{van~de Hulst}},
  \emph{\bibinfo{title}{Light Scattering by Small Particles}}
  (\bibinfo{publisher}{John Wiley \& Sons, Inc.}, \bibinfo{address}{New York},
  \bibinfo{year}{1957}).

\bibitem[{\citenamefont{Boardman and Paranjape}(1977)}]{Boardman:1977a}
\bibinfo{author}{\bibfnamefont{A.~D.} \bibnamefont{Boardman}} \bibnamefont{and}
  \bibinfo{author}{\bibfnamefont{B.~V.} \bibnamefont{Paranjape}},
  \bibinfo{journal}{J. Phys. F: Metal Phys.} \textbf{\bibinfo{volume}{7}},
  \bibinfo{pages}{1935} (\bibinfo{year}{1977}).

\end{thebibliography}
\end{document}